# Plasma-Activated Water and Phenolic Compounds: A Potent Combined Strategy Against Yeast Resilience

Bernard Gitura Kimani[a], Ramin Mehrabifard[a], Oleksandr Galmiz[a], and Zdenko Machala[a*]

[a]Division of Environmental Physics, Faculty of Mathematics, Physics and Informatics, Comenius University in Bratislava, Mlynská dolina, 842 48 Bratislava, Slovakia

corresponding author: machala@fmph.uniba.sk

## Abstract

Effective fungal control is essential to prevent spoilage, contamination, and infections in the realms of food, healthcare, and industry. This study explores the antifungal activity of plasma-activated water (PAW) generated by transient spark (TS) electrical discharge combined with natural phenolics bioactive compounds on the planktonic growth, biofilm formation, and surface adhesion of yeast *Saccharomyces cerevisiae* and *Wickerhamomyces anomalous*. The phenolics were assayed at 1 or 2 mg/mL for planktonic growth after determining their minimum inhibitory concentration. The biofilms were grown on sterile microscope slides for 48 h, stained with acridine orange or calcofluor stain, and observed under a fluorescence microscope. PAW was used immediately after plasma treatment of tap water to create PAW-phenolic solutions. The combination of PAW with cinnamic acid and vanillin showed the most significant antiyeast activity against planktonic and biofilm growth, as well as surface adhesion. The biofilms formed with the PAW-phenolic combination were notably fragmented, and residual cells were structurally damaged. A complete inhibition of biofilm formation was observed when 500 µg/mL of cinnamic acid was used in combination with PAW. The findings indicate a strong potential of combining PAW with phenolic compounds that represents an effective novel strategy for preventing fungal growth.

## 1. Introduction

Yeasts are versatile microorganisms useful in food production, biotechnology, and the pharmaceutical industry, contributing to the production of a variety of products [1]. However, uncontrolled fungal growth can lead to harmful outcomes, including opportunistic infections in healthcare environments [2] and spoilage in food and beverage industries [3]. Contaminating yeasts such as *Dekkera* sp. are known to cause wine spoilage [3]. Although chemical fungicides and heat treatments remain common control strategies, their use is increasingly constrained by concerns about resistance development, environmental toxicity, and adverse effects on product quality, creating demand for more sustainable antifungal approaches [1, 4]. The combination of plasma-activated water (PAW) with natural bioactive compounds is an exciting new paradigm in fungal control that may offer an effective and environmentally friendly alternative to classical antifungals [5].

Cold atmospheric plasma (CAP) has attracted considerable attention in recent years as an effective non-thermal approach for microbial control [6]. In food processing, where thermal treatments may impair sensory and nutritional properties [7], CAP offers the advantage of operating under ambient conditions while generating reactive oxygen and nitrogen species (RONS) that can damage microbial cells through oxidative mechanisms [8]. The efficacy of CAP treatment on microbial cells depends on plasma characteristics, plasma dose, and on the susceptibility of the target microorganism [8]. CAP systems can be constituted by several types of electrical discharges, including dielectric barrier discharge (DBD), plasma jets, corona and spark discharges, and gliding arc discharges, each with distinct properties and application ranges [8, 9, 10]. Different CAP sources can be used in agriculture to promote seed germination, control pests, and decontaminate agricultural produce, and in biomedical contexts for wound healing and the sterilization of biomedical devices [10]. Owing to its tunable chemistry, versatility, and relatively low environmental footprint, CAP is viewed as a promising platform for sustainable fungal decontamination in food, agricultural, and biomedical settings [9].

PAW has emerged as a promising, environmentally friendly antimicrobial solution with significant potential in food safety and clinical sanitation [10]. Unlike direct plasma treatments, PAW offers distinct advantages in biomedical and food applications by mitigating risks associated with electric currents, thermal tissue injury, and UV radiation exposure. The antimicrobial efficacy of PAW is primarily driven by the enrichment of RONS during the exposure of water or aqueous solutions to CAP. This process introduces highly transient species, such as hydroxyl radicals ($\cdot OH$) and nitric oxide (NO), alongside more stable, long-lived species including hydrogen peroxide ($H_2O_2$), nitrite ($NO_2^-$), nitrate ($NO_3^-$), and ozone ($O_3$) [11]. The composition of PAW, and consequently its potential applications, depends strongly on the type of plasma source and the working gas, as well as the energy delivered during the treatment [10,11]. During plasma generation in atmospheric air, plasma-formed RONS are transported into the liquid. Disparity exists between liquid phase-RONS and gas phase-RONS due to the differences in solubility and the chemical reactions that occur in the liquid phase [10, 11]. RONS transported into PAW can cause oxidative and nitrosative stress on fungal cells, which can disrupt membrane lipid layers, denature proteins and modify nucleic acids [8]. Various physicochemical factors such as the pH of the solution and storage conditions can affect the reactivity of PAW and its antifungal activity [12]. Taken together, these factors highlight the challenge of optimizing PAW performance, raising important questions about its consistency and efficacy beyond the laboratory scale. A key transition challenge in industrial implementation is developing an integrated system for large scale PAW production and maintaining PAW effectiveness and reliability when moving from controlled laboratory conditions to the variable environments of large-scale production. Several studies have demonstrated the scalability of PAW, including pilot-scale fresh-cut lettuce processing with up to 5-log microbial reduction [13], an industrial system with 10-100 L/h PAW production capacity and over 8-log microbial disinfection efficiency [14], and an off-site plasma device producing 2 L/h PAW with > 99.9% bactericidal efficiency [15].

Natural bioactive compounds have both fungicidal and fungistatic effects; however, their effectiveness may be hindered by issues of stability and bioavailability [16]. Blending these compounds with well-established antifungal agents or novel delivery methods can immensely increase efficacy and spectrum of activity [17]. These compounds exert their antifungal effects through changing plasma membrane lipid profile, apoptosis induction, producing reactive oxygen species (ROS), and interfering with key fungal signaling pathways [18, 19]. They come in different structural classes and are synthesized through the shikimate and phenylpropanoid biosynthetic pathways [20]. These compounds can insert into microbial lipid bilayers, modulate membrane fluidity, bind essential transition metal ions, and block fungal enzymes [21, 22, 23]. Due to their structural diversity, biodegradability, low toxicity to the environment, and low chance of causing resistance, natural phenolics are becoming crucial in developing eco-friendly and sustainable antifungal strategies.

Although PAW and natural phenolics have each been studied independently for their antiyeast properties, their combined application remains largely unexplored, representing a significant knowledge gap. Such combinations may enhance antifungal activity through complementary mechanisms [17, 24], including membrane permeabilization, induction of fungal apoptosis, organelle disruption, or the generation of cytotoxic chemical byproducts against fungal cells [17, 24, 25]. This combined approach could improve antifungal efficacy, broaden the range of susceptible fungal targets, and reduce the likelihood of antimicrobial resistance development [17].

In this study, the antifungal potential of PAW generated by a 1 kHz TS plasma discharge is investigated, in combination with selected natural phenolics, namely cinnamic acid, vanillin, gallic acid, and *p*-coumaric acid, against yeast *S. cerevisiae* SZMC 1279 and *W. anomalous* SZMC 8061Mo. The PAW-phenolic combinations were evaluated for their ability to inhibit planktonic growth, biofilm formation, and surface adhesion, and were compared with PAW alone, phenolics alone, and untreated controls. The structural integrity of yeast biofilms with and without PAW-phenolic treatment was further assessed using fluorescence microscopy. Our findings provide strong evidence for the enhanced antiyeast efficacy of PAW-phenolic formulations and highlight their potential as a sustainable strategy for fungal control.

## 2. Materials and Methods

### *2.1 Transient spark (TS) discharge*

The TS discharge was generated by a stainless-steel power electrode connected to a high voltage DC power supply through a 10 MΩ resistor (Fig. 1) [10]. The power electrode had a diameter of 0.80 mm, was 40 mm long, and possessed a bevel tip. A metal ring used as a grounded electrode was submerged at the bottom of the Petri dish in sterile tap water (Fig. 1). The TS discharge treatment parameters were set at a duration of 10 min for a 10 mL volume of sterile tap water, with a steady ratio of 1 min/mL for optimal surface area to volume ratio, consistent treatment conditions, and enough sample recovery for later physicochemical and antimicrobial tests. The 10-minute exposure time per 10 mL was chosen based on initial tests that compared different treatment times. Times shorter than 10 minutes produced less RONS, leading to lower anti-yeast effects (Fig. S1), while longer discharge times induced water acidification and had higher energy consumption without corresponding antiyeast effects (Fig. S1). The gap between the tip of the power electrode and water surface was 10 mm and the frequency was 1 kHz. The Petri dish containing sterile tap water had an internal diameter of 48 mm and a height of 14 mm. A high voltage probe (Tektronix, TDS 2024C) was used to measure the voltage at the power electrode. The discharge current for TS plasma was measured using Rogowski current monitor Pearson electronics 2877 (1 V/1 A). The TS set-up and the Current-Voltage waveform of the TS discharge is shown in Fig. 1A and 1B, respectively.

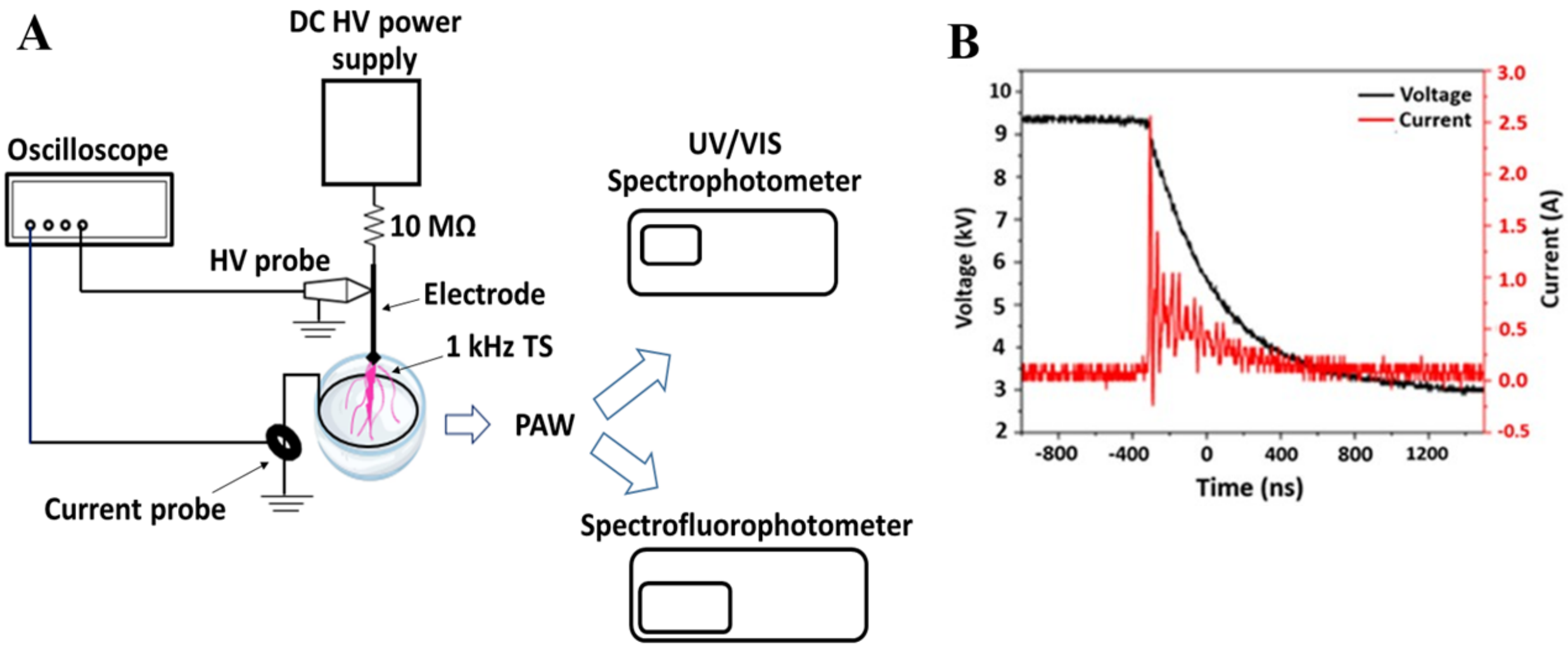

**Fig. 1**. (**A**) Schematic diagram of the experimental setup used for PAW generation by 1 kHz TS discharge, (**B**) characteristic voltage and current waveforms of TS discharge.

Generally, the measurement of plasma power through voltage and frequency involves the calculation of the power delivered to the plasma based on its electrical characteristics. The instantaneous power, $P(t)$, at any specific moment ($t$) is defined as (1):

$$P(t) = V(t).I(t) \qquad (1)$$

Where $V(t)$ is the instantaneous voltage, and $I(t)$ is the instantaneous current (1). To find the average power over one TS pulse cycle, we integrate $P(t)$ over the period ($T$) determined by the frequency $f$ (2):

$$P_{avg} = \frac{1}{T}\int_0^T P(t)dt = f \int_0^T V(t)I(t)\ dt \qquad (2)$$

*2.2 PAW analysis*

The main plasma-generated aqueous species, which include $H_2O_2$, $NO_2^-$, and $NO_3^-$, were identified through UV-Vis absorption spectroscopy (Shimadzu UV-1800). $H_2O_2$ was quantified using the titanium oxysulfate ($TiOSO_4$) assay under acidic conditions, which relies on the formation of yellow pertitanic acid ($H_2TiO_4$) [26, 27, 28]. To stabilize the samples immediately after plasma discharge and prevent $H_2O_2$ decomposition via reaction with $NO_2^-$, 60 mM sodium azide solution ($NaN_3$) was added. The absorbance of the resulting complex was measured at 407 nm with UV-Vis spectrophotometer.

The determination of dissolved $O_3$ concentration was performed using the indigo trisulfonate decolourization assay, which is particularly specific to detect $O_3$ because of its oxidative degradation of the indigo blue dye [26, 29]. Under acidic conditions, $O_3$ decolorizes indigo potassium trisulfonate, leading to the formation of the colourless compound isatin. The decrease in absorbance at 600 nm ($\varepsilon = 2.38 \times 10^4$ $M^{-1}$ $cm^{-1}$) correlates linearly with the increasing concentration of dissolved $O_3$.

$NO_2^-$ levels were measured using the Griess reagent assay [26, 30]. $NO_2^-$ react with the Griess reagent to form a deep purple azo compound, which exhibits an absorption maximum at 540 nm. We utilized 2,6-dimethylphenol (DMP) (Sigma-Aldrich, USA) to measure $NO_3^-$ in PAW. DMP interacts with $NO_3^-$ but also $NO_2^-$ in sulfuric and phosphoric acid solutions to generate 4-nitro-2,6-dimethylphenol, which is detectable by spectrophotometry at about 330 nm [31]. The calibration curve is used to determine the $NOx^-$ concentration in PAW, which is proportional to this absorbance. $NO_3^-$ concentration was subsequently calculated by subtracting the previously determined $NO_2^-$ concentration from the total $NOx^-$ value [26, 30]. The concentration of the ·OH was quantified using chemical dosimetry with terephthalic acid [32] leading to the formation of 2-hydroxyterephthalic acid (HTA), which was detected through fluorescence spectrometry at excitation and emission wavelengths of 310 nm and 425 nm, respectively [33].

*2.3 PAW parameters measurement*

The pH testing was conducted using a calibrated pH meter (WTW 3110, Weilheim, Germany). For measuring temperature and electrical conductivity (EC), a digital conductivity meter (GMH 3430, GREISINGER electronic, Germany) was employed. A 4-in-1 multimeter (Noyafa Digital, Philippines) was utilized to measure total dissolved solids (TDS).

*2.4 Yeast Strains and Growth Conditions*

*Saccharomyces cerevisiae* SZMC 1279 and *Wickerhamomyces anomalous* SZMC 8061Mo were used in the assays. The yeast was cultivated on a malt extract medium composed of 5 g/L yeast extract, 5 g/L glucose, and 50 mL/L of 20% (v/v) malt extract [34]. Fresh yeast cultures were prepared prior to each assay by incubating for 24 hours at 30 °C. Following serial dilution, cell numbers were determined by counting in a Bürker chamber under a light microscope [34].

*2.5 Phenolic compounds*

To ensure representation of different classes of phenolic compounds, cinnamic acid and *p*-coumaric acid (hydroxycinnamic acids), gallic acid (hydroxybenzoic acid), and vanillin (phenolic aldehyde) were selected for the assay (Table 1). These compounds were obtained from Sigma-Aldrich (Munich, Germany). All the selected phenolics had demonstrated their antiyeast activity in previous studies [1, 34]. The phenolic compounds were tested at concentrations of 1 or 2 mg/mL in the planktonic assay. These concentrations were selected after preliminary screening (Fig. S2) to identify balanced and effective working levels for comparative analysis and were consistent with the minimum inhibitory concentration (MIC) values previously determined and confirmed in the present study [1, 34]. These concentrations fall within the limits allowed for use as food additives; moreover, all the compounds are naturally present in foods and considered safe in normal dietary intake.

**Table 1**

Phenolic compounds used in the study.

| Compound | Chemical structure |
| --- | --- |
| Cinnamic acid | |
| Vanillin | |
| *p*-coumaric acid | |
| Gallic acid | |

*2.6 PAW-phenolic treatment of planktonic yeast cells*

Two approaches were used to evaluate the effects of PAW-phenolic treatments on planktonic yeasts. (1) In the indirect treatment, PAW was generated by a 1 kHz TS discharge for 10 min and used immediately to dissolve phenolic compounds to prepare stock solutions; 100 µL of the PAW-phenolic solution was then mixed with 100 µL of yeast suspension prepared in double-strength sterile media to obtain final phenolic concentrations of 2 or 1 mg/mL and a yeast concentration of $10^6$ colony-forming units/mL (CFU/mL), followed by static incubation at 30 °C for 24 h. (2) In the direct treatment, a 10 mL yeast suspension in sterile tap water was exposed to TS discharge for 10 min, after which 100 µL of the treated suspension was combined with 100 µL of phenolic compounds prepared in double-strength sterile medium to yield the same final phenolic and yeast concentrations as in indirect treatment, and incubated under similar conditions. The effects of PAW-only and phenolics-only on planktonic yeast growth were also assessed. Treatment efficacy was determined by serial dilution and plating on agar, and CFUs were enumerated after 24 h of incubation at 30 °C. Positive controls consisted of untreated yeast mixed with sterile medium (final concentration = $10^6$ CFU/mL), while negative controls contained sterile medium and PAW-phenolics without yeast inoculation for sterility test. Yeast growth in PAW-phenolics, PAW-only, and phenolics-only was compared with the untreated control, and growth inhibition was expressed as the difference relative to the control growth. All measurements were carried out in three biological and three technical parallels.

*2.7 Biofilm formation inhibition*

The effect of PAW-phenolic combinations on biofilm formation was evaluated based on our previous methods with a slight modification [34]. Considering that sessile cells in biofilms exhibit different physiological states compared to planktonic cells and reduced antimicrobial penetration and concentration gradients, especially due to the protective effects of extracellular polymeric substances (EPS), lower phenolic concentrations (125 - 500 µg/mL) were screened to investigate anti-biofilm formation effects without entirely inhibiting growth, with 500 µg/mL showing consistent inhibitory effects for the assayed compounds (cinnamic acid and vanillin). Wells of a 96-well polystyrene microtiter plate (Sarstedt, Nümbrecht, Germany) were inoculated with 200 µL of a 24 h yeast culture with approximately $10^8$ CFU/mL concentration subjected to a 1 kHz TS discharge treatment for 10 minutes. The plate was incubated at 30 °C for 4 hours to allow initial yeast adhesion. After adhesion, the planktonic non-adherent cells were gently aspirated from each well. Wells were then rinsed with sterile physiological saline to remove residual non-adherent cells. The plate was left to air-dry for 10 minutes in a laminar flow cabinet. Subsequently, 200 µL of phenolic compounds (at a final concentration of 500 µg/mL) prepared in a 1:1 (v/v) mixture of PAW and double-strength sterile growth medium were added to each well. Positive control wells received 200 µL of standard growth medium. Negative control wells had no adhered cells; they received the PAW + phenolic solution in the growth medium for sterility test. For PAW-only treatments, 100 µL of PAW was mixed with an equal volume of double-strength sterile medium (1:1, v/v) and added to wells containing adhered yeast cells previously exposed to TS discharge. For phenolic-only treatments, a 500 µg/mL solution of the phenolic compound was prepared in sterile medium and 200 µL of this solution was added to wells containing adhered yeast cells that had not been exposed to TS discharge. The plates were incubated for 48 hours at 30 °C to allow biofilm development. After incubation, planktonic cells were removed by gentle aspiration, and the wells were washed twice with sterile physiological saline to eliminate loosely attached cells. To harvest the biofilms, 200 µL of sterile physiological saline was added to each well. The biofilms were dislodged through a series of pipetting, mild mechanical agitation, and 2 min of ultrasonication at 20 kHz, with the integrity of the cells verified under microscopic observation [35]. The resulting cell suspensions were transferred to sterile microcentrifuge tubes, gently vortexed, serially diluted, and plated on agar. CFUs were enumerated after 24 hours of incubation at 30 °C. Each experiment was performed in triplicate with three technical replicates.

*2.8 Fluorescence Microscopy Analysis of Biofilms*

Yeast biofilms were developed on sterile microscope glass slides (26 × 76 mm) placed in the centre of sterile 90 mm Petri dishes. Yeast cells were cultured in sterile growth medium at 30 °C for 24 h, followed by gentle homogenization via vortexing. A 10 mL aliquot of the suspension (approximately $10^8$ CFU/mL) was subjected to 10 min TS discharge, 8 mL of the TS treated yeast cells was added to each Petri dish to cover the glass slide then incubated statically at 30 °C for 4 h to facilitate initial adhesion. After this period, planktonic cells were removed by pipetting, and the slides were gently rinsed with sterile physiological saline to eliminate non-adherent cells. PAW was generated using a 1 kHz-10 min TS and used to prepare stock solutions of phenolic compounds. These were then diluted 1:1 with double-strength growth medium to obtain a final concentration of 500 µg/mL for either vanillin or cinnamic acid. A total of 8 mL of the PAW-phenolic treatment solution was applied to each Petri dish containing the TS treated adhered cells on microscope glass slides. For PAW-only treatments, 4 mL of PAW was mixed with an equal volume of double-strength sterile medium (1:1, v/v) and applied to Petri dishes containing microscope slides with adhered yeast cells previously exposed to TS discharge. For phenolic-only treatments, a 500 μg/mL solution of the phenolic compound was prepared in sterile medium, and 8 mL of this solution was added to Petri dishes containing microscope slides with adhered yeast cells that had not been exposed to TS discharge. Positive control samples were prepared by adding 8 mL of sterile media on adhered cells not exposed to TS discharge in the absence of PAW-phenolic combinations, while negative control had PAW-phenolic in slides without yeast cells for sterility test. All samples were incubated at 30 °C for 48 h to allow biofilm development. Following incubation, biofilms were stained with 20 μL of 1% (w/v) acridine orange or calcofluor solution (Sigma-Aldrich, Munich, Germany), then gently rinsed with distilled water to remove excess stain after 5 min incubation. Biofilm visualization was performed using a Keyence BZ-X Series fluorescence microscope at 100× magnification.

### *2.9 Yeast Adhesion Analysis*

The anti-adhesion activity of PAW-phenolic (vanillin or cinnamic acid) was assessed using a 96-well microtiter plate assay adapted from our previous methods with minor modifications [34]. Yeast overnight cultures were prepared as described in Section 2.8. A 10-mL aliquot of the 24-h suspension was exposed to TS discharge (1 kHz, 10 min). Phenolic stock solutions were prepared in double-strength culture medium and mixed with TS discharge-treated yeast cells (1:1, v/v) to yield 500 μg/mL phenolic and $10^8$ CFU/mL yeast. Aliquots (200 μL) were transferred into wells and statically incubated at 30 °C for 4 h to allow adhesion. For PAW-only treatments, 100 μL of TS discharge-treated cells was combined with an equal volume of double-strength culture medium, to a final cell density as in PAW-phenolic treatments. Phenolic-only treatments consisted of phenolic solution in inoculated growth medium (500 μg/mL, $10^8$ CFU/mL). Growth medium inoculated with yeasts ($10^8$ CFU/mL) without PAW or phenolic served as the positive control, while PAW-phenolic mixtures without inoculum served as negative controls for sterility test. After incubation, non-adherent cells were gently aspirated, and wells were washed with sterile physiological saline. Adherent cells were recovered by adding 200 μL of physiological saline per well, mixing with pipette, mild mechanical agitation and transferring the suspension to sterile microcentrifuge tubes, vortexing, serially diluting in sterile deionized water, and plating on agar [35, 36]. Plates were incubated at 30 °C for 24 h, and CFUs were enumerated. All experiments were performed in triplicate.

### *2.10 Spectral analysis of phenolic compounds*

The UV-Vis spectral characteristics of cinnamic acid, vanillin, *p*-coumaric acid, and gallic acid were analyzed by UV-Vis spectrophotometry. This analysis was used as a rapid screening method to detect PAW-induced changes in the absorption spectral behavior of natural phenolics. The solutions of all the compounds at concentrations of 1-10 μg/mL were prepared in sterile tap water and in PAW. Absorbance spectra of all compounds were measured at room temperature using a Shimadzu UV-1800 UV-Vis spectrophotometer at the resolution of 1 nm in the wavelength range of 190-700 nm.

### *2.11 Statistical Analysis*

Assays were performed in at least three independent experiments and data were expressed as means ± standard deviation. Basic statistical analysis of data, such as calculation of means and standard deviations, was conducted using Microsoft Office Excel 2016 function. Significance was calculated by one-way analysis of variance (ANOVA) followed by Tukey's multiple comparison test in the GraphPad Prism 6.00 software (GraphPad Software Inc., San Diego, CA, USA). A *p*-value of $<0.05$ was considered as statistically significant.

## 3. Results and discussion

### *3.1. TS discharge*

In this study, TS discharge was used as the main plasma source for generating PAW. TS is a self-pulsing, DC-driven discharge that periodically shifts from a streamer to short-lived, high-current spark pulses [28]. This process generates high-density RONS while the generated plasma stays in a non-thermal state [28]. We chose this plasma treatment for its better energy efficiency and chemical reactivity compared to other discharge types, based on our previous research [10, 28, Machala 2019]. For example, glow discharges have high reactivity but require much more power, while diffuse streamers usually produce lower RONS concentrations [37]. Additionally, the TS discharge mode offers more stable water activation than atmospheric pressure plasma jets (APPJ), which are very sensitive to duty cycles [38]. Therefore, in this investigation, we aimed to use TS discharge to enable a significant transfer of bioactive species into the liquid, which is necessary for antifungal effectiveness when combined with natural compounds.

### *3.2 PAW analysis*

The physicochemical characteristics of water, both prior to and following the TS discharge treatment, are presented in Table 2. The pH of sterile tap water reduced from 7.9 to 7.5 after TS discharge treatment. The EC and TDS rose by 36 $\mu S$/cm and 30 ppm, respectively, whereas the water temperature had a minimal increase of 0.8 °C post plasma treatment. The power utilized by the TS discharge, as indicated by the voltage-current (V-I) plot, was 4 W.

**Table 2**

Physicochemical parameters of water before (control) and after plasma treatment.

| | pH | EC ($\mu S$/cm) | TDS (ppm) | Temperature (°C) | Power (W) |
|---|---|---|---|---|---|
| Control | $7.9 \pm 0.2$ | $390 \pm 37.0$ | $215 \pm 18.0$ | $23 \pm 1.0$ | - |
| 10 min TS | $7.5 \pm 0.2$ | $426 \pm 34.0$ | $245 \pm 17.0$ | $23.8 \pm 0.6$ | $4 \pm 0.5$ |

The slight pH reduction from 7.9 to 7.5 following TS treatment is consistent with the dissolution of RONS in tap water [39]. Since tap water contains buffering ions such as bicarbonates and carbonates, the pH reduction was not as significant as typical in plasma-treated distilled water where the absence of buffering capacity allows for a greater acidification [39]. This shows how liquid/water composition affects the scale of the pH change after plasma treatment. The increase in EC and TDS by 36 $\mu S$/cm and 30 ppm, respectively, demonstrates production and dissolution of ionic species during the plasma treatment [40]. Tap water itself have ions such as calcium, magnesium, and chloride, etc. RONS adds to the ionic strength further. The resulting rise in EC and TDS confirms the enrichment of water with reactive ionic solutes [40]. The slight temperature increase (0.8 °C) after 10 min of PAW production, confirmed the non-thermal character of the TS discharge. The low-heating effect shows that physicochemical transformation is due to reactive chemistry rather than bulk thermal effects, this is important for treating heat-sensitive substrates such as fresh vegetables and fruits, or biomedical materials [41].

The concentration of RONS was measured immediately after plasma treatment. Results are presented in Table 3. $O_3$ had the lowest concentration at $6.01 \pm 0.8$ μM, followed by ·OH at a concentration of 15 ± 2 μM. $H_2O_2$ was measured at 398 ± 15 μM. The highest concentration was that of $NO_2^-$ and $NO_3^-$: 1327 ± 137 μM and 1250 ± 190 μM, respectively, indicating robust nitrogen oxidation processes upon TS discharge exposure. Previous studies in our research group have shown that in dry air, TS generates high concentrations of NO and $NO_2$, increasing linearly with increasing input energy density [10].

While freshly prepared PAW is an effective antimicrobial agent, using it directly on biological samples requires caution due to possible oxidative side effects [42]. High concentration of RONS can induce lipid peroxidation and create reactive compounds that can negatively impact biological materials [42]. In addition, organic matter may lessen the PAW's effectiveness, resulting in inconsistent outcomes [43]. Therefore, it is crucial to set up optimized, substrate-specific conditions and perform thorough testing to ensure safe application and effectiveness. Once these conditions are established, attention must shift to avoiding exposure of microbial populations to sublethal PAW concentrations [44]. Such conditions may trigger environmental stress responses, enhancing cellular resilience [44]. Thus, the PAW/plasma treatment conditions should be carefully optimized to prevent stress adaptation of microorganisms.

**Table 3**

Concentration of RONS in PAW after 10 minutes' treatment by TS plasma

| $O_3$ ($\mu M$) | ·OH ($\mu M$) | $H_2O_2$ ($\mu M$) | $NO_2^-$ ($\mu M$) | $NO_3^-$ ($\mu M$) |
|---|---|---|---|---|
| 6.01 ± 0.8 | 15 ± 2 | 398 ± 15 | 1327 ± 137 | 1250 ± 190 |

*3.3. PAW-phenolic treatment of planktonic yeasts*

We evaluated the antiyeast efficacy of combining PAW with natural phenolic compounds against planktonic *S. cerevisiae* and *W. anomalous* in both indirect and direct treatment modalities. The phenolics screened included cinnamic acid, vanillin, *p*-coumaric acid, and gallic acid at concentrations of 1 or 2 mg/mL. Significant differences in anti-yeast activity were observed depending on both the treatment mode and the phenolic compound used. For the indirect treatment, cinnamic acid exhibited the strongest inhibitory effect, with 2 mg/mL of cinnamic acid in PAW leading to 8.2 log inhibition in *S. cerevisiae* (Fig. 2A) and 8.3 log inhibition in *W. anomalous* (Fig. 3A). At 2 mg/mL, cinnamic acid alone had 3.9 log inhibition in *S. cerevisiae* (Fig. 2A) and 4.5 log inhibition in *W. anomalous* (Fig. 3A). At 1 mg/mL, PAW-cinnamic acid had a 4 log growth inhibition in *S. cerevisiae*, which was significantly higher than in PAW or cinnamic acid alone, which had 1.6 and 1.9 log inhibition, respectively (Fig. 2A). In *W. anomalous*, PAW-cinnamic acid had a 4.8 log inhibition at 1 mg/mL, while cinnamic acid and PAW alone had 3.6 and 1.3 log growth inhibition, respectively (Fig. 3A). While PAW and cinnamic acid individually showed limited inhibitory effects, their combination resulted in a significant increase in bioactivity. The higher bioactivity of PAW-cinnamic acid is probably a result of the oxidative transformation of cinnamic acid by PAW-RONS (Fig. S3), which created molecules with higher antiyeast effects. This transformation is supported by the presence of a broader and more intense absorption band in the range of 230-300 nm in the UV-Vis spectra (Fig. S3). The reduced absorbance of the PAW-cinnamic acid between 200-220 nm may further indicate a partial modification of the native molecule. However, these spectral changes only provide indirect evidence, and the exact structures of the modified products are yet to be elucidated. Future studies that include detailed chemical characterization are necessary to understand the nature of the transformation and the role of the derivatives in the observed improved anti-yeast effects.

PAW-vanillin combination also had enhanced anti-yeast activity, with 2 mg/mL of vanillin combined with PAW having a 4 log growth inhibition in *S. cerevisiae* (Fig. 2B) and 4.6 log growth inhibition in *W. anomalous* (Fig. 3B). 1 mg/mL of vanillin with PAW resulted in a 3 log growth inhibition in *S. cerevisiae* (Fig. 2B) and 3.6 log inhibition in *W. anomalous*, which was higher than in vanillin or PAW alone. In *S. cerevisiae,* vanillin alone had 3.2 and 1.9 log growth inhibition at 2 mg/mL and 1 mg/mL, respectively (Fig. 2B), while in *W. anomalous*, it had 3.6 and 2.5 log inhibition at 2 mg/mL and 1 mg/mL, respectively (Fig. 3B).

The UV-Vis absorption spectra showed a pronounced increase in absorbance for PAW-vanillin, especially within the 260-380 nm range (Fig. S4), together with changes in spectral shape when compared with vanillin prepared in tap water (Fig. S4). This implies that plasma activation modifies the chemical environment and/or structure of vanillin, shifting it towards a more biologically active state.

In contrast, *p*-coumaric acid exhibited moderate antiyeast activity when combined with PAW, resulting in 2.9 and 1.9 log inhibition at concentrations of 2 and 1 mg/mL, respectively in *S. cerevisiae* (Fig. 2C), and 2.4 and 1.5 log inhibition at 2 and 1 mg/mL, respectively in *W. anomalous* (Fig. 3C). These combined effects were lower than the expected additive outcomes based on the individual performances of each treatment in the two yeasts. In *S. cerevisiae*, *p*-coumaric acid alone had 2.1 and 0.8 log inhibition at 2 and 1 mg/mL, respectively (Fig. 2C). In *W. anomalous,* 2 and 1 mg/mL of *p*-coumaric acid alone had a 1.6 and 0.8 log inhibition, respectively. In *p*-coumaric acid, only a limited increase in anti-yeast activity was observed for the combined treatment relative to the individual treatments. The slightly higher bioactivity of PAW-*p*-coumaric acid may be due to partial oxidative modification of *p*-coumaric acid by PAW-RONS (Fig. S5), resulting in the formation of derivatives with slightly enhanced antiyeast activity.

Structurally, the presence of the -OH group in the para position distinguishes *p*-coumaric acid from the unsubstituted cinnamic acid (Table 1), which exhibits greater activity, likely due to its decreased radical-scavenging capacity and enhanced permeability through membranes [45].

Gallic acid, a strong antioxidant, exhibited the weakest antiyeast activity in combination with PAW, with no significant enhancement observed over individual treatments (Fig. 2D and 3D). The UV-Vis analysis revealed noticeable differences between the gallic acid solutions prepared in sterile tap water and PAW (Fig. S6A). Gallic acid in sterile tap water displayed the characteristic absorption at 260-270 nm (Fig. S6A). On the other hand, PAW-treated gallic acid showed reduced UV absorbance between 260-270 nm with an asymmetric band around 340-370 nm (Fig. S6A), possibly signifying the presence of oxidation products [46, 47]. The colour transformation from colourless for the gallic acid prepared in tap water to yellow for the PAW-treated gallic acid solution (Fig. S6B), further supports the likely formation of oxidized gallic acid derivatives [46], with low bioactivity against yeasts.

In the direct treatment of planktonic cells, yeast suspensions were first exposed to TS discharge for 10 minutes and then treated with phenolics. PAW-cinnamic acid had a complete growth inhibition at both 1 and 2 mg/mL of cinnamic concentration for both *S. cerevisiae* and *W. anomalous* (Fig. 4). This superior performance in direct treatment is likely due to the simultaneous physical and chemical impact of plasma discharge. We hypothesize that these plasma-induced physical changes increase cell wall permeability, facilitating the entry of cinnamic acid into the cell [48]. Future studies should verify the proposed permeability changes. In addition, direct TS treatment exposed the yeast cells to both short-lived and long-lived species, direct ultraviolet radiation, transient electric fields, and acoustic waves. These factors may further compromise cellular membranes and facilitate the uptake of phenolics. PAW-driven oxidative transformation of the native cinnamic acid (Fig. S3) may enable multi-target antifungal mechanisms.

PAW-vanillin also demonstrated higher antiyeast inhibition when compared to individual treatments. In *S. cerevisiae*, PAW-vanillin had a complete inhibition at 2 mg/mL and a 3.2 log inhibition at 1 mg/mL (Fig. 4A). For *W. anomalous*, the PAW-vanillin log inhibitions were 5.4 and 4.3 at 2 and 1 mg/mL of vanillin concentration, respectively (Fig. 4B). PAW-induced modifications on vanillin might positively influence its antiyeast activity [49]. PAW-*p*-coumaric acid had a moderate anti-yeast effect for both yeasts (Fig. 4). However, the combined effect was greater than either treatment used on its own (Fig. 4). The limited solubility in water, poor stability, low permeability across membranes and antioxidant character of *p*-coumaric acid could impede its ability to fully take advantage of plasma-induced damage of yeasts [50]. Just like in indirect treatment, PAW-gallic acid had the weakest antiyeast activity of the combined compounds (Fig. 4), probably by scavenging critical RONS and the formation of oxidized gallic acid derivatives with low antiyeast activity (Fig. S6) [46].

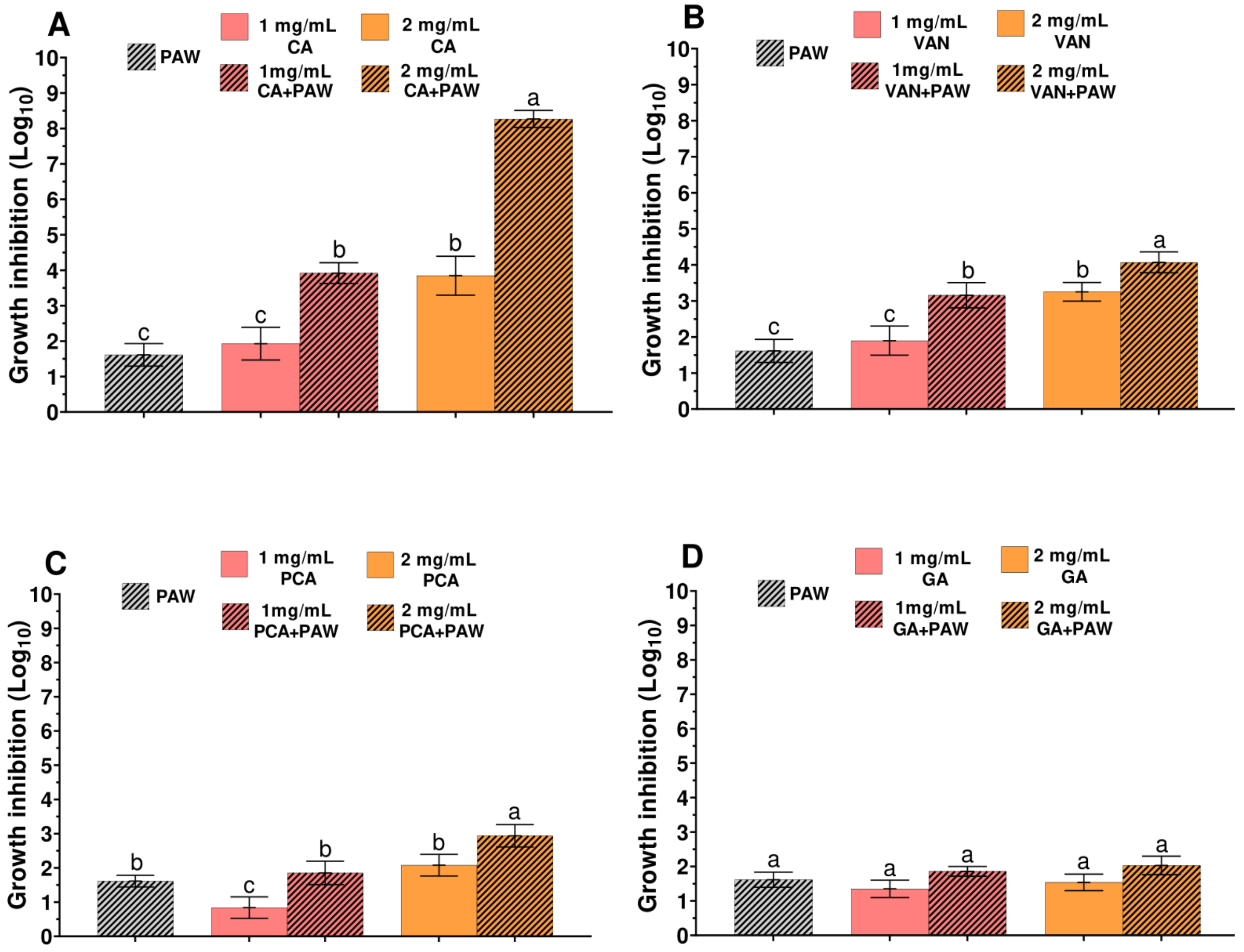

**Fig. 2**. Growth inhibition of planktonic *S. cerevisiae* SZMC 1279 following treatment with PAW and phenolic compounds. Growth inhibition ($\log_{10}$) was assessed after exposure to PAW alone or in combination with phenolic compounds: (A) cinnamic acid (CA), (B) vanillin (VAN), (C) *p*-coumaric acid (PCA), and (D) gallic acid (GA), each applied at 1 mg/mL or 2 mg/mL. Bars represent mean growth inhibition compared to untreated controls. Error bars indicate standard deviations. Different lowercase letters above the bars denote statistically significant differences among treatments as determined by ANOVA followed by Tukey's post hoc test ($p < 0.05$).

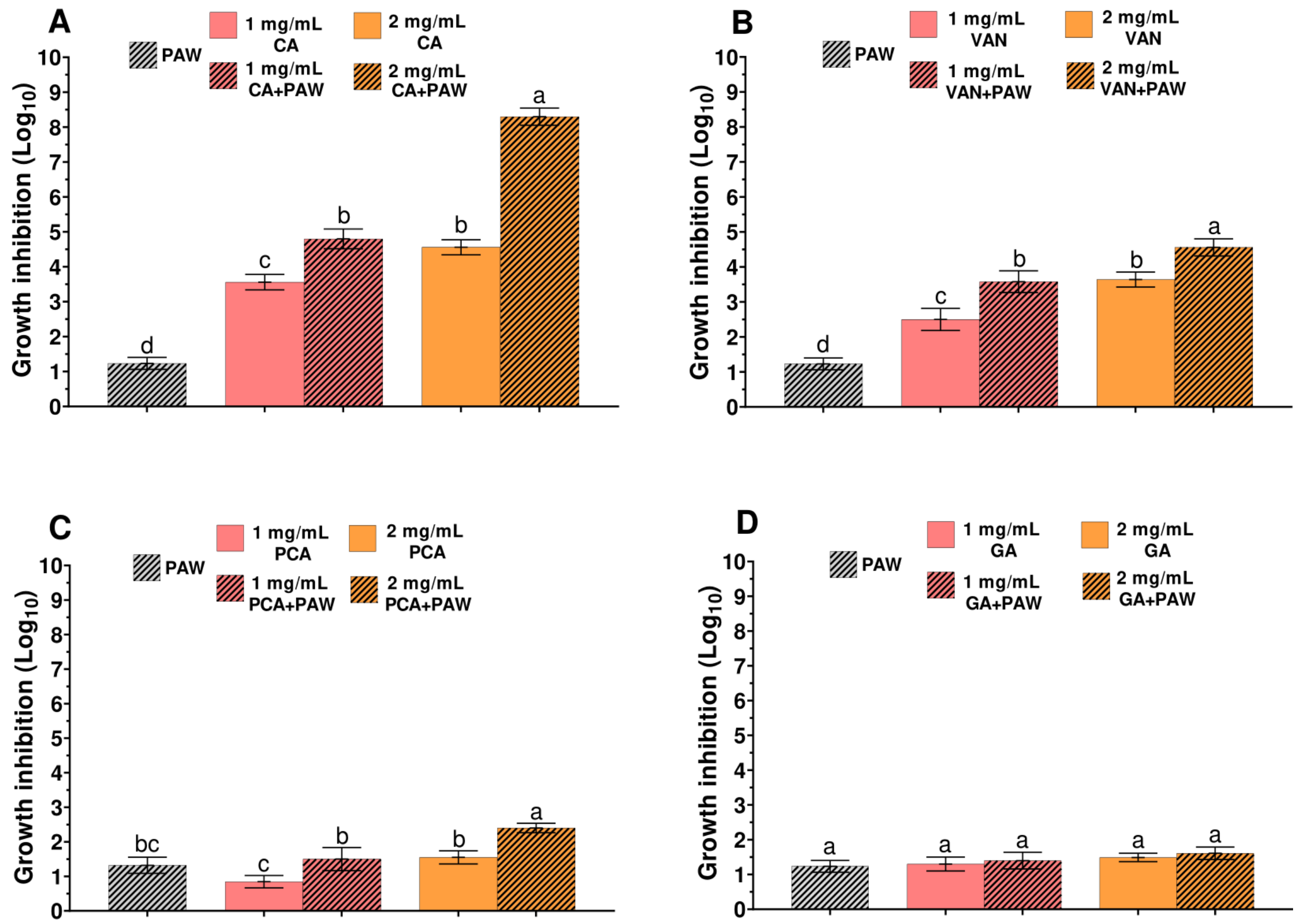

**Fig. 3**. Growth inhibition of planktonic *W. anomalous* SZMC 8061Mo following treatment with PAW and phenolic compounds. Growth inhibition ($log_{10}$) was assessed after exposure to PAW alone or in combination with phenolic compounds: (A) cinnamic acid (CA), (B) vanillin (VAN), (C) *p*-coumaric acid (PCA), and (D) gallic acid (GA), each applied at 1 mg/mL or 2 mg/mL. Bars represent mean growth inhibition compared to untreated controls. Error bars indicate standard deviations. Different lowercase letters above the bars denote statistically significant differences among treatments as determined by ANOVA followed by Tukey's post hoc test ($p < 0.05$).

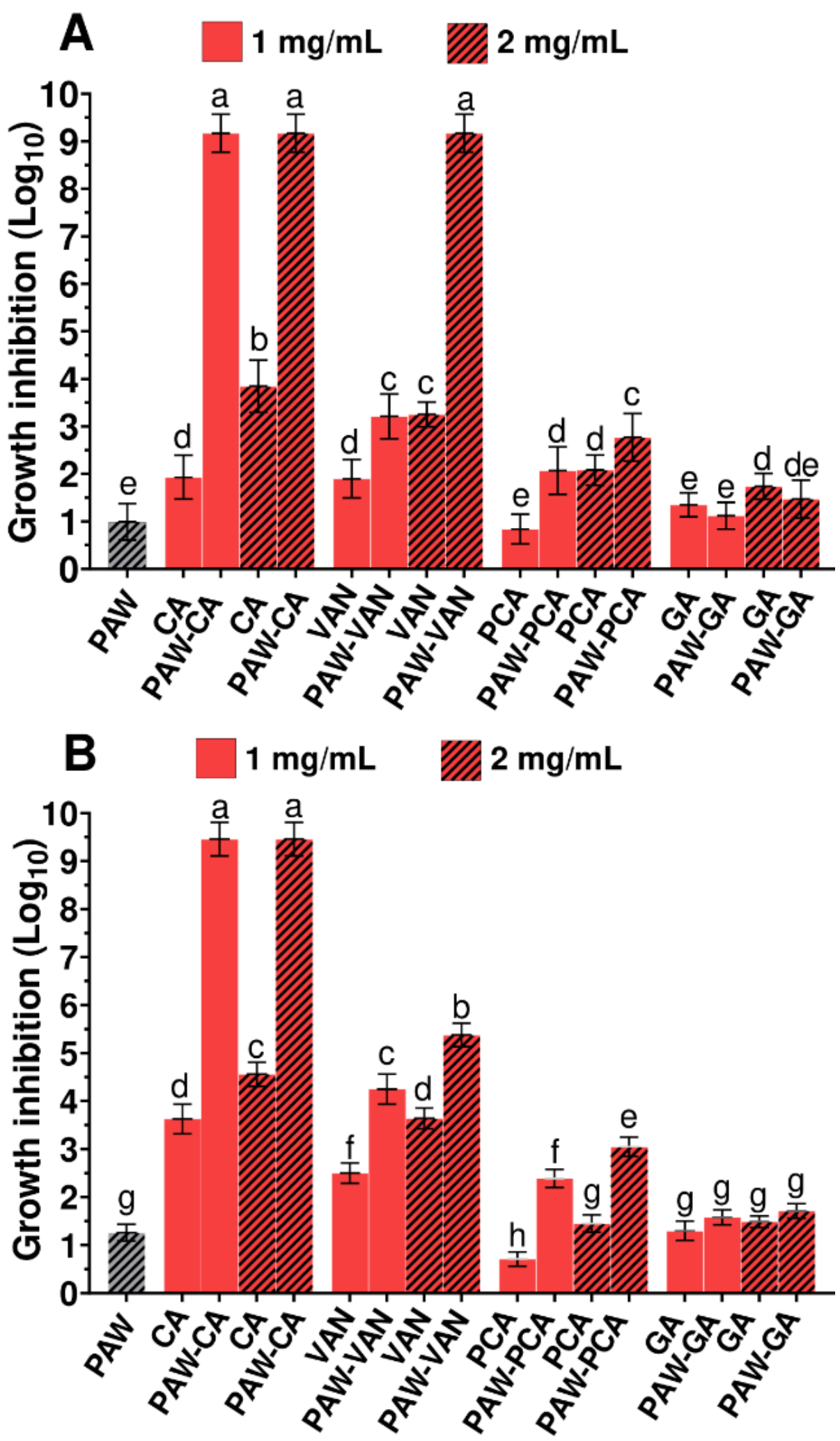

**Fig. 4.** Growth inhibition of planktonic *S. cerevisiae* SZMC 1279 (A), and *W. anomalous* SZMC 8061Mo (B), following direct treatment with TS discharge to generate PAW *in situ*. Yeast cells were subsequently exposed to cinnamic acid (CA), vanillin (VAN), *p*-coumaric acid (PCA), or gallic acid (GA), applied individually or in combination with PAW at 1 mg/mL or 2 mg/mL. Growth inhibition is expressed as $\log_{10}$ reductions relative to untreated controls. Error bars represent standard deviations. Different letters indicate statistically significant differences (one-way ANOVA with Tukey's test, $p < 0.05$).

### *3.4 Biofilm formation inhibition*

Biofilm formation was assessed using a standard 96-well polystyrene microtiter plate assay. Two phenolic compounds - cinnamic acid and vanillin - were selected for evaluation at a concentration of 500 μg/mL, based on their previously observed efficacy in inhibiting planktonic cells when combined with PAW. The combination of PAW and cinnamic acid resulted in a complete biofilm inhibition in both yeasts (Fig. 5), demonstrating a potent combined effect that far exceeds the capabilities of either treatment alone. While PAW, when used alone, resulted in a modest 1.3 log inhibition in *S. cerevisiae* (Fig. 5A) and 1 log in *W. anomalus* (Fig. 5B), its contribution in the combination seems to be crucial, likely priming the biofilm for deeper penetration. PAW-vanillin was equally effective in *S. cerevisiae* with the combination leading to a 7 log biofilm inhibition, while vanillin alone had a 3.5 log biofilm inhibition (Fig. 5A). Lower antibiofilm activity was observed in *W. anomalous*, with PAW-vanillin having a 4.3 log inhibition and vanillin alone having a 2.9 log inhibition (Fig. 5B). The increased activity of these combinations can be attributed to the diverse assault of PAW-RONS on the physical barriers of the biofilm which potentiates the phenolics. By potentially improving the permeability of microbial membranes and inhibiting the development of EPS, PAW may enhance the antibiofilm properties of phenolic compounds by easing the diffusion limitations that the biofilm matrix usually creates [51, 52, 53]. Additionally, the chemical activation of cinnamic acid and vanillin through PAW-RONS, as illustrated in Fig. S3 and S4, respectively, may have generated bioactive molecular species, thereby potentially strengthening their inhibitory interactions with both the protective matrix and the embedded microbial cells. Future studies should validate these proposed antibiofilm mechanisms.

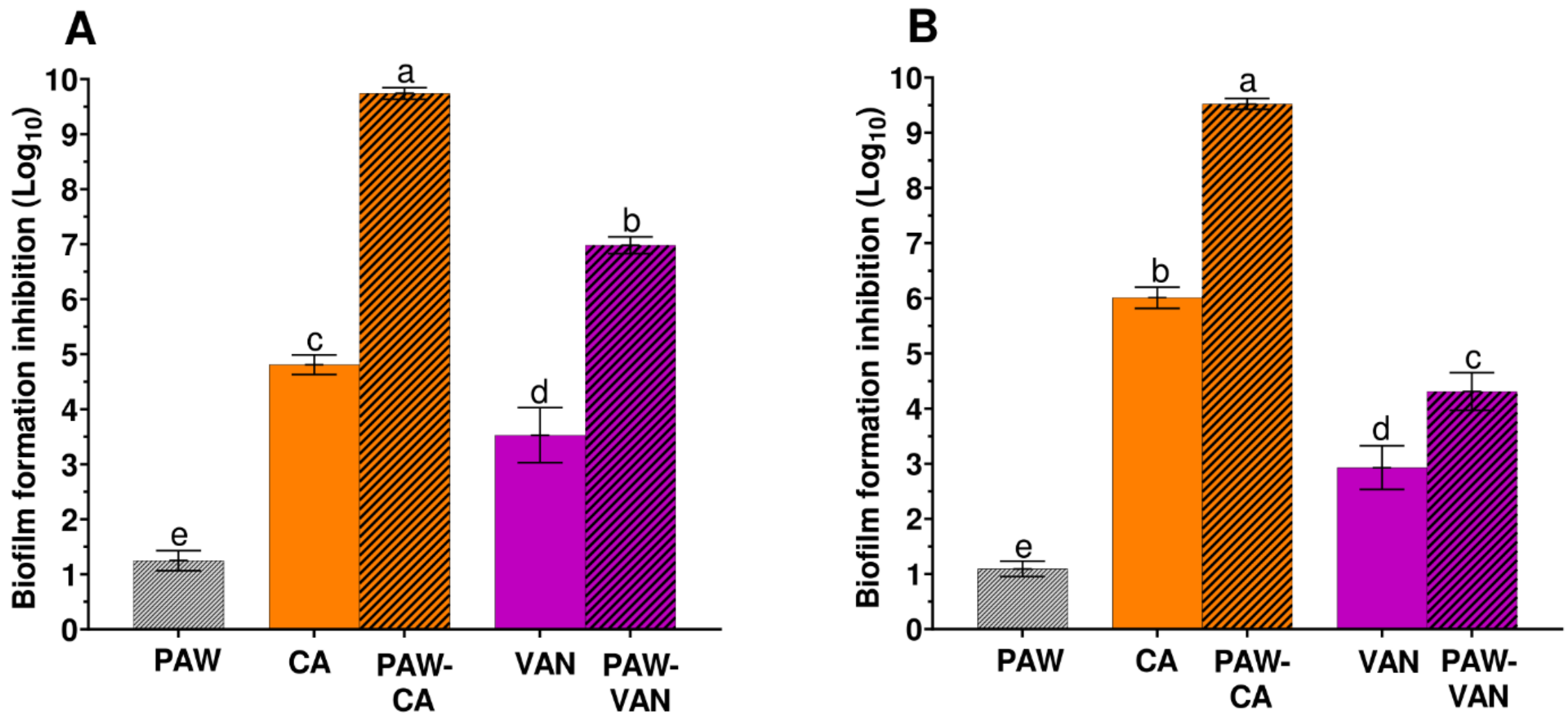

**Fig. 5.** Effect of different treatments on the biofilm formation of (A) *S. cerevisiae* SZMC 1279 and (B) *W. anomalous* SZMC 8061Mo. Treatments included PAW, cinnamic acid (CA), vanillin (VAN), and phenolics combinations with PAW (PAW-CA, PAW-VAN). 500 µg/mL of phenolics was used in PAW-phenolic treatment. The results shown represent the $log_{10}$ reduction in biofilm formation compared to the untreated control. Error bars indicate standard deviation. Different letters above the bars denote statistically significant differences among the treatments as determined by one-way ANOVA followed by Tukey's multiple comparison test ($p < 0.05$).

*3.5 Fluorescence Microscopy Analysis of Biofilms*

To assess structural alterations in yeast biofilms following treatment, fluorescence microscopy was performed on biofilms developed on sterile microscope glass slides under different conditions as described in section 2.8. Microscopic visualization revealed that untreated samples had a dense, biofilm network (Fig. 6). The partial reduction in biofilm density observed with individual treatments (Fig. 6I) demonstrates that individual agents lack the potency to fully suppress the transition from initial attachment to consolidated, stable biofilm architecture. In contrast, samples subjected to the PAW-phenolic combination exhibited pronounced ultrastructural degradation (Fig. 6). The sparse distribution of residual yeast cell fragments and the presence of diffused nucleic acids directly indicate extensive membrane permeabilization and biofilm matrix destabilization specifically triggered by the PAW-cinnamic acid combination. This is further supported by the calcofluor white fluorescence signal, which showed significant impairment of the yeast cell wall (Fig. 6II). The transition from the uniform, oval shapes of control cells to the irregular, shrunken morphologies in the PAW-cinnamic acid group provides visual evidence of reduced turgor pressure and structural collapse. These observations confirm that the combined treatment led to a terminal failure of the cell wall and plasma membrane, resulting in the leakage of cytoplasmic materials.

To validate the loss of cellular viability, post-treatment regrowth assays were performed. Yeast cells recovered from the PAW-cinnamic acid treatment group failed to grow on solid agar medium, confirming that the structural damage induced by PAW-cinnamic was lethal and irreversible. The combination of PAW and phenolics may influence the lipid homeostasis of yeast, potentially leading to increased membrane rigidity and a higher susceptibility to permeabilization (Fig. 6II) [54]. The consequent destabilization of membrane integrity likely promoted pore formation, allowing the efflux of cytosolic components (Fig. 6II) [55]. In addition, the PAW-phenolics treatment may enhance lipid peroxidation through reactive oxygen species-mediated processes [56]. This oxidative damage could extend beyond membrane lipids to include peroxidation of proteins and enzymes involved in the biosynthesis of EPS and capsular polysaccharides [57, 58]. Inhibition or degradation of these components would lead to a marked reduction in EPS volume and yeast cell size, as can be seen in PAW-phenolic treated cells in Figure 6. These cumulative effects would manifest as increased cell surface roughness, and diminished biofilm architectural integrity, ultimately reducing the biofilm's protective function and persistence. Although these results strongly suggest that PAW-phenolics disrupt lipid homeostasis, promote oxidative damage, and impair EPS biosynthesis as the central mechanisms of biofilm inhibition, more studies are required to validate these proposed interactions and confirm the mechanistic basis of PAW-phenolics-mediated biofilm disruption.

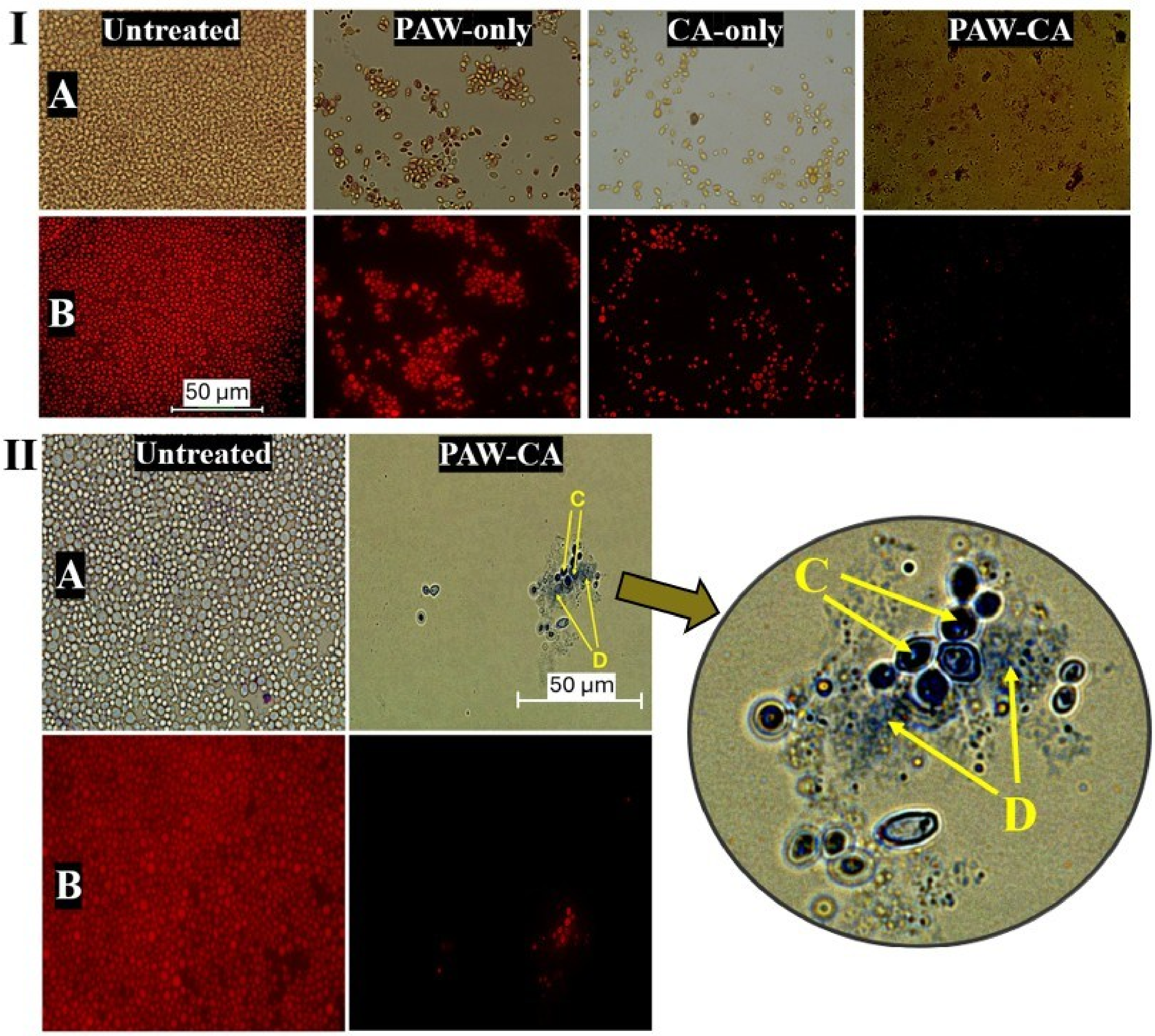

**Fig. 6.** Biofilm formation of *S. cerevisiae* SZMC 1279 (I) and *W. anomalous* SZMC 8061Mo (II) under different treatment conditions. Cinnamic acid (CA) was applied at a concentration of 500 µg/mL. After 48 hours of incubation, the *S. cerevisiae* biofilm was stained with acridine orange and *W. anomalous* with calcofluor stain and visualized using a fluorescence microscope at 100× magnification. The untreated images (control) represent biofilm formation without PAW,

cinnamic acid or their combination (PAW-CA). Rows IA and IIA, show bright-field images, and IB and IIB show corresponding fluorescence images. Arrow **C** shows residual yeast cells, while **D** denotes cytoplasmic remnants and cell debris.

### *3.6 Yeast Adhesion Analysis*

Investigations of inhibition of yeasts' adherence to non-biological surfaces, including plastic polymers and glass, are few, indicating a significant gap in the literature. The adhesion assay demonstrated notable variations in the ability of the two yeast strains to adhere to abiotic surfaces after treatment with PAW, phenolic compounds, or their combinations. *S. cerevisiae* was notably resistant to the combined treatment compared to *W. anomalous*, with PAW-cinnamic acid having a 2.8 log inhibition in *S. cerevisiae* (Fig. 7A) and 6 log inhibition in *W. anomalous* (Fig. 7B). In *S. cerevisiae*, cinnamic acid and PAW alone had 1.2 and 1.9 log inhibition (Fig. 7A), respectively, while in *W. anomalous*, cinnamic acid and PAW alone had 3.1 and 1.5 adhesion inhibition, respectively (Fig. 7B). PAW-vanillin had a lower inhibition effect in the two yeast strains with the combination having a 2 log inhibition in *S. cerevisiae* (Fig. 7A), and a 4 log inhibition in *W. anomalous* (Fig. 7B). Vanillin alone had a 0.7 and 2.3 log inhibition in *S. cerevisiae* and *W. anomalous*, respectively (Fig. 7). This pronounced difference in susceptibility can be directly attributed to the distinct cell wall compositions of the two species. *S. cerevisiae* possesses heavily glycosylated adhesins within a dense β-glucan-chitin matrix, which has been associated with greater tolerance to physical and chemical stresses and may differ from the cell surface organization of non-conventional yeasts such as *W. anomalus* [59]**.** Adhesion is an essential survival process in yeasts, enabling transitions between unicellular and multicellular forms [60, 61]. By interfering with these initial attachment stages, the PAW-phenolic approach disrupts the transition from unicellular to multicellular forms, thereby limiting colonization of abiotic surfaces.

The adhesion process is mainly governed by FLO gene-encoded adhesins, whose expression is regulated by nutrient and stress-responsive pathways and further modulated through epigenetic mechanisms [61]. This system enables yeasts to survive in fluctuating conditions. However, this survival trait in yeasts poses challenges in different sectors. In agriculture, fungal adhesion causes fungal diseases and post-harvest spoilage, and in medical settings, it promotes biofilm formation on medical devices, e.g. catheters, and recurrent infections. In this context, the use of PAW-phenolics approach to control initial yeast adhesion on abiotic surfaces presents an environmentally safe method in microbial management in agricultural, industrial, and biomedical applications [62].

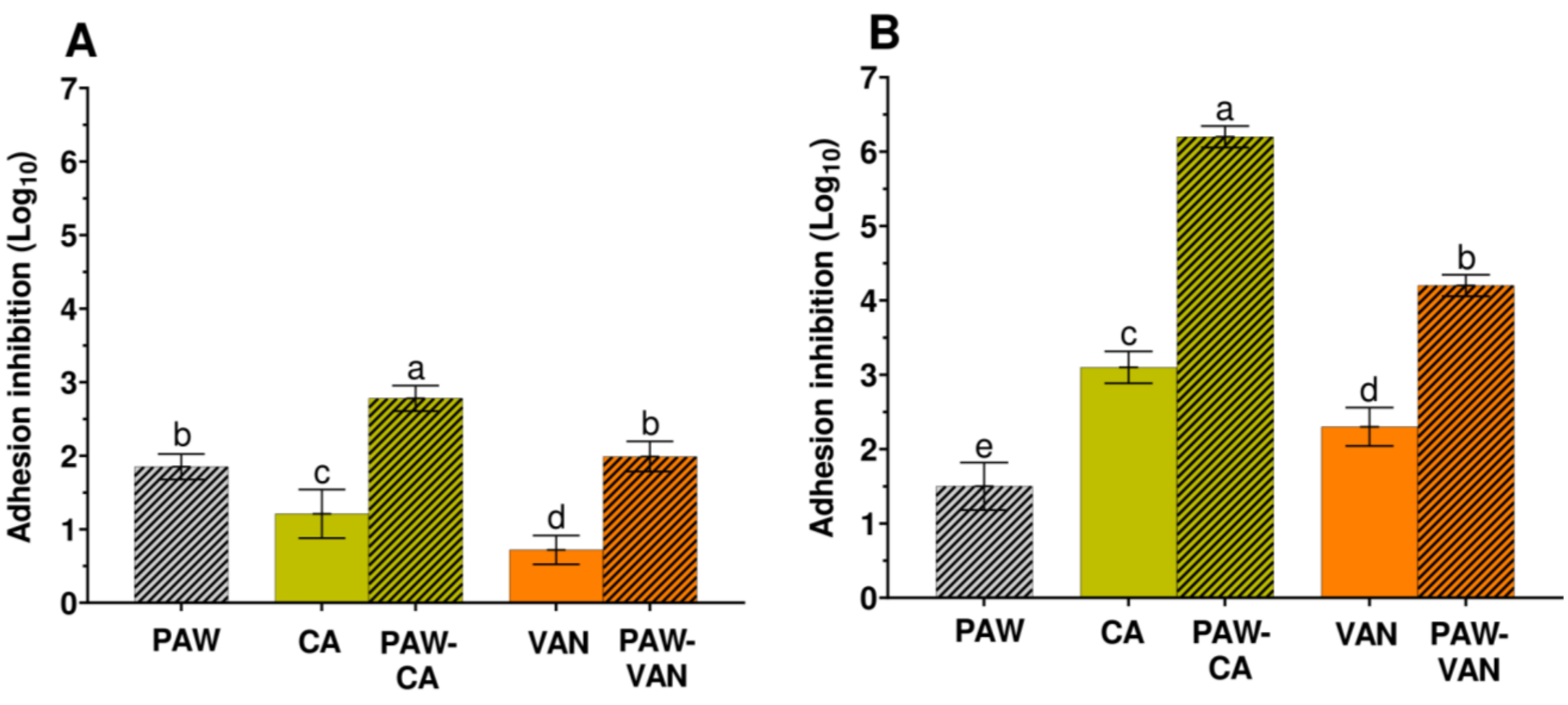

**Fig. 7.** Effect of different treatments on adhesion to a polystyrene surface of (A) *S. cerevisiae* SZMC 1279 and (B) *W. anomalous* SZMC 8061Mo. Treatments included PAW, cinnamic acid (CA), vanillin (VAN), and phenolics combinations with PAW (PAW-CA, PAW-VAN). A total of 500 μg/mL of phenolics was used in the treatments. The results represent the $\log_{10}$ reduction in adhesion compared to the untreated control. Error bars indicate standard deviation. Different letters

above the bars indicate statistically significant differences among the treatments, as determined by one-way ANOVA followed by Tukey's multiple comparison test ($p < 0.05$).

## 4. Conclusion

The present study investigates antifungal activity by a novel approach of combining PAW with natural phenolic compounds on the planktonic growth, biofilm formation, and surface adhesion of yeast *Saccharomyces cerevisiae* and *Wickerhamomyces anomalous*. The combination of PAW with cinnamic acid at a concentration of 1 and 2 mg/mL during direct treatment led to a total inhibition of planktonic growth in both yeast species, whereas a lower concentration of 500 µg/mL of cinnamic acid combined with PAW effectively prevented their biofilm formation. From our results, it is evident that the effectiveness of PAW combined with phenolics depends on whether PAW is generated in the presence of yeasts (for direct treatment), or in their absence (for indirect treatment), with the direct treatment proving to be more effective than the indirect treatment.

In addition, the chemical nature of the selected phenolic compounds significantly contributes to the antiyeast activity. PAW combined with cinnamic acid showed the highest antiyeast activity, most likely due to its chemical properties, membrane-permeabilizing capacity, and PAW-driven oxidative transformation of native cinnamic acid, which may enable multi-target antifungal mechanisms. However, these mechanistic interpretations remain hypothetical and require direct experimental verification in future studies. Vanillin also showed strong, though slightly less pronounced activity when combined with PAW. On the other hand, *p*-coumaric acid and gallic acid showed moderate to minimal effects. The improved antifungal effects between PAW and selected phenolics were likely influenced by mechanistic factors such as the disruption of membranes, impairment of the yeast cell wall, oxidative stress induction, and destabilization of the EPS (in biofilms). Based on our findings, the choice of phenolic compound is an important step in maximizing the antifungal effects of cold plasma and PAW treatments, favouring compounds with low antioxidant activity. In addition, this study reveals that the antifungal effects of PAW-phenolics are strain-specific; therefore, it is crucial to consider species and strain differences when developing PAW-phenolic antifungal strategies. It appears that the main role of PAW in improving the bioactivity of the combinations is to deliver RONS that disrupt the integrity of yeast cell membranes and affect intracellular redox balance. They also hinder the formation and stability of EPS in biofilms and induce oxidative changes in phenolic compounds, thereby making them more biologically active. This process enhances the penetration and antifungal activity of the phenolics. Future studies should confirm the proposed mechanisms.

In summary, PAW-phenolic combinations showed a strongly enhanced antifungal activity against selected yeast. Despite this increased antiyeast activity observed in laboratory settings, further validation in real-world applications is necessary to determine their stability and practical effectiveness over time. Future studies should also elucidate the molecular mechanisms that enhance fungal activity for certain PAW-phenolic combinations and the impact of PAW-RONS on the chemical nature and functionality of the phenolic compounds.

**Acknowledgment**

Funded by the EU NextGenerationEU through the Recovery and Resilience Plan for Slovakia under the project No. 09I03-03-V03-00033, 09I03-03-V05-00012, 09I03-03-V04-00094, and Slovak Research and Development Agency APVV-22-0247.

Dr. Miklos Takó provided the yeasts from the University of Szeged Microbiological Collection (SZMC).

**Conflict of interest**

Authors claim no conflicting interests.